\documentclass[aps,english,showpacs,twocolumn,prl]{revtex4-1}
\usepackage{amssymb}
\usepackage{amsmath}
\usepackage{graphicx}
\usepackage{epsfig}
\usepackage{color,xcolor}

\begin{document}

\title{Untying links through anti-parity-time-symmetric coupling}
\author{H. C. Wu}
\author{X. M. Yang}
\author{L. Jin}
\email{jinliang@nankai.edu.cn}
\author{Z. Song}
\email{songtc@nankai.edu.cn}
\affiliation{School of Physics, Nankai University, Tianjin 300071, China}

\begin{abstract}
We reveal how the vector field links are untied under the influence of
anti-parity-time-symmetric couplings in a dissipative sublattice-symmetric
topological photonic crystal lattice. The topology of the
quasi-one-dimensional two-band system is encoded in the geometric topology
of the vector fields associated with the Bloch Hamiltonian. The linked
vector fields reflect the topology of the nontrivial phase. The topological
phase transition occurs concomitantly with the untying of the vector field
link at the exceptional points. Counterintuitively, more dissipation
constructively creates a nontrivial topology. The linking number predicts
the number of topological photonic zero modes.
\end{abstract}

\maketitle

\textit{Introduction.}---The monopole, skyrmion, and vortex are robust
topological structures that have attracted much research interests in
optics, quantum physics, and condensed matter physics. The zero-intensity
lines of light field, in the form of isolated knotted nodal lines, are
observed as optical vortices in experiments \cite{Dennis10,IrvinePRL13}. The
creation, observation, and investigation on the static and dynamical
properties of quantum knots in Bose-Einstein condensates are reported \cite%
{Ueda08,Hall16,Hall19}. Topological phases are extremely stable due to
topological protection, even in non-Hermitian topological systems \cite%
{Hasan,XLQi,YFChen,DWZhang,Cooper,Ozawa,YXu,LonghiPRBSR,Hirsbrunner,SLZhu20,DWZhang20A,CHLee20}%
. In topological systems, zero-energy Fermi surfaces can form knotted or
linked nodal lines \cite%
{SCZhang,JMHou,GChang,Ezawa,ZWang,ZWang2,Ackerman,DLD,Carlstrom,Bergholtz,JPHu19L}%
. Alternatively, the fictitious magnetic field of a topological system with
topological defects, associated with vortex or anti-vortex textures,
reflects nontrivial topology \cite{SLin,PXue19}. The fictitious magnetic
field is a geometrical analogy of the magnetic field. A nontrivial topology
can be extracted from the spin polarization, which is associated with a
two-band topological system and defines a vector field. The vector field
exhibits a link or knot topology in the topologically nontrivial phase. We
visualize the vortex links, present alterations to the vector field topology
under the influence of anti-parity-time-symmetric (anti-$\mathcal{PT}$%
-symmetric) couplings, and reveal the untying of vector field links
associated with a topological phase transition. Counterintuitively, we
demonstrate that more dissipation constructively creates nontrivial topology.

\textit{Dissipative photonic crystal lattice.---}Photonic crystal is an
excellent platform for the study of topological physics \cite{Hafezi}. The
non-Hermitian quasicrystal \cite{Longhi19L,YXu20B}, high-order topological
phase \cite{XLuo19,TLiu19,CHLeePRL19,Ezawa19}, robust edge state \cite%
{Esaki,Schomerus,JLPRA,Weimann,MPan,Poli,ZKL1905,XDZhang}, topological
lasing \cite%
{Feng,Conti,LonghiAnnPhys,Segev,Khajavikhan,Skryabin,Carusotto,Jean,Parto,Iwamoto}
in photonic crystals were discovered. Here, we consider a photonic crystal
lattice of coupled resonators, as schematically illustrated in Fig.~\ref%
{fig1}(a). The colored resonators are the primary resonators, and the white
and gray resonators are the auxiliary resonators, evanescently coupled to
the primary resonators. The primary resonators have an identical resonant
frequency, $\omega _{c}$. The auxiliary resonators induce the effective
couplings between primary resonators. The blue, red, and white resonators
are passive; the gray resonators have dissipation. The dissipation rate is
much larger than the coupling strength between the gray auxiliary resonators
and the primary resonators. The large dissipation enables the adiabatical
elimination of the auxiliary resonator light field and results in a
reciprocal anti-$\mathcal{PT}$-symmetric coupling $-i\gamma $ between two
primary resonators; in the on-resonance condition, the strength $\gamma $
equals the product of two evanescent couplings between the gray auxiliary
resonator and its two adjacent primary resonators divided by the dissipation
rate of the gray auxiliary resonator \cite{LYou}. The other two effective
coupling strengths induced by the white auxiliary resonators are denoted as $%
J$ and $\kappa $.

\begin{figure}[b]
\includegraphics[ bb=60 365 485 630, width=8.8 cm,clip]{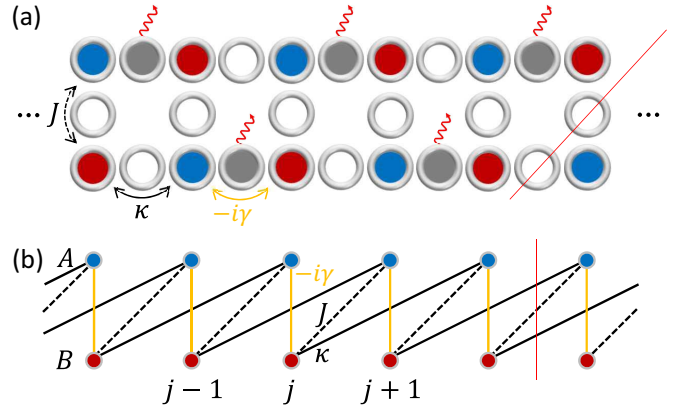}
\caption{(a) Quasi-1D coupled resonator array. The primary resonators (in
blue and red) are effectively coupled by the auxiliary resonators (in gray
and white). (b) Schematic of (a). The open boundary of (b) corresponds to
the sideling cut of (a), indicated by the solid red lines.}
\label{fig1}
\end{figure}

\begin{figure*}[tb]
\includegraphics[ bb=10 0 585 190, width=17.8 cm,clip]{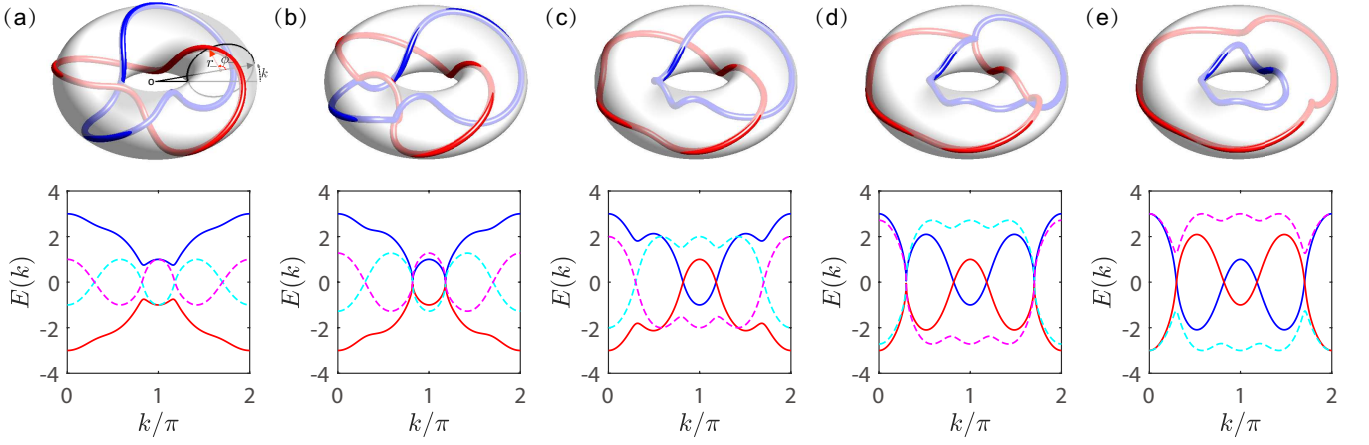}
\caption{Vector field curves $\mathbf{F}_{\pm}\left( k\right) $ and energy
bands for distinct topological phases. (a) $\protect\gamma =1$, (b) $\protect%
\gamma =1.274$, (c) $\protect\gamma =2$, (d) $\protect\gamma =2.715$, and
(e) $\protect\gamma =3$. The linking numbers are $L=-2$, $-1$, and $0$ for
(a), (c), and (e); (b) and (d) are the gapless phases. Other parameters are $%
J=1$ and $\protect\kappa =2$. The energy bands are fully complex except at
the crossing points of their imaginary parts.}
\label{fig2}
\end{figure*}

In coupled mode theory \cite{CMT}, the equations of motion for the coupled
resonator array are%
\begin{eqnarray}
i\frac{\mathrm{d}}{\mathrm{d}t}\psi _{A,j} &=&\omega _{c}\psi _{A,j}-i\gamma
\psi _{B,j}+J\psi _{B,j-1}+\kappa \psi _{B,j-2}, \\
i\frac{\mathrm{d}}{\mathrm{d}t}\psi _{B,j} &=&\omega _{c}\psi _{B,j}-i\gamma
\psi _{A,j}+J\psi _{A,j+1}+\kappa \psi _{A,j+2},
\end{eqnarray}%
where $\psi _{A,j}$ and $\psi _{B,j}$ are the wavefunction amplitudes for
the corresponding sites $j$ of the sublattices $A$ and $B$. The schematic of
the quasi-one-dimensional (quasi-1D) bipartite lattice is shown in Fig. \ref%
{fig1}(b).

Applying the Fourier transformation, the equations of motion for the
eigenmode with momentum $k$ reduce to%
\begin{eqnarray}
i\frac{\mathrm{d}}{\mathrm{d}t}\psi _{A,k} &=&\omega _{c}\psi _{A,k}+\left(
-i\gamma +Je^{-ik}+\kappa e^{-2ik}\right) \psi _{B,k}, \\
i\frac{\mathrm{d}}{\mathrm{d}t}\psi _{B,k} &=&\omega _{c}\psi _{B,k}+\left(
-i\gamma +Je^{ik}+\kappa e^{2ik}\right) \psi _{A,k}.
\end{eqnarray}%
After the common resonant frequency $\omega _{c}$ is removed for both
sublattices, the two-band system is described by the Bloch Hamiltonian $%
h_{k}=(\vec{d}-i\vec{\Gamma})\cdot \vec{\sigma}$, where $\vec{\sigma}=\left(
\sigma _{x},\sigma _{y},\sigma _{z}\right) $ is the Pauli matrix for spin-$%
1/2$ and $\vec{d}-i\vec{\Gamma}$ is the fictitious magnetic field including
the real part $\vec{d}=\left( J\cos k+\kappa \cos \left( 2k\right) ,J\sin
k+\kappa \sin \left( 2k\right) ,0\right) $ and the imaginary part $\vec{%
\Gamma}=(\gamma ,0,0)$ \cite{XZhang20L,Hofmann20}. The two energy bands are 
\begin{equation}
\varepsilon _{k}^{\pm }=\pm \lbrack (d_{x}-i\gamma )^{2}+d_{y}^{2}]^{-1/2}.
\end{equation}%
The corresponding eigenmodes are%
\begin{equation}
|\psi _{k}^{\pm }\rangle =[1,(d_{x}-i\gamma +id_{y})/\varepsilon _{k}^{\pm
}]^{T}/\mathcal{N},
\end{equation}%
which are normalized to unity $\langle \psi _{k}^{\pm }|\psi _{k}^{\pm
}\rangle =1$. The average values of Pauli matrices under the eigenmode yield
the spin polarization \cite{SLin} and define a momentum dependent \textit{%
unity} 3D vector field $\mathbf{F}_{\pm }\left( k\right) =\left( F_{\pm
,x}\left( k\right) ,F_{\pm ,y}\left( k\right) ,F_{\pm ,z}\left( k\right)
\right) $. All of the three components $F_{\pm ,x,y,z}\left( k\right)
=\langle \psi _{k}^{\pm }|\sigma _{x,y,z}|\psi _{k}^{\pm }\rangle $ are
gauge independent periodic functions of the momentum $k$ and can be
expressed in the form 
\begin{equation}
\mathbf{F}_{\pm }\left( k\right) =\left( \sin \theta _{\pm }\cos \phi _{\pm
},\sin \theta _{\pm }\sin \phi _{\pm },\cos \theta _{\pm }\right) .
\end{equation}

\textit{Linking topology}.\textit{---}In geometry, 1D topology describes 1D
closed curves in 3D space described by knot theory \cite{KnotTheory}. Here,
the vector field information is encoded in $\left( \theta _{\pm },\phi _{\pm
}\right) $, and Fig.~\ref{fig2} shows the vector fields $\mathbf{F}%
_{+}\left( k\right) $ and $\mathbf{F}_{-}\left( k\right) $ in different
topological phases; they are mapped onto a torus as two closed curves, and
their interrelation provides a visualization of the band topology of $h_{k}$%
: The linked vector field curves represent the nontrivial topology. In Fig.~%
\ref{fig2}, the torus has a major radius of $R_{0}$, which is the distance
from the center of the tube to the center of the torus. $r_{0}$ is the minor
radius of the tube. $k$ is taken as the toroidal direction, and $\mathbf{R}%
_{\pm }(k)=(R_{0}+r_{\pm }\cos \phi _{\pm })\vec{e}_{r}+r_{\pm }\sin \phi
_{\pm }\vec{e}_{z}$; $r_{\pm }=r_{0}\sin \theta _{\pm }$ is selected to
ensure that $\mathbf{R}_{\pm }(k)$ forms a closed curve. The curve detaches
from the surface of the torus when $\left\vert r_{\pm }\right\vert \neq
r_{0} $. The braiding of the two vector field curves is characterized by the
linking number%
\begin{equation}
L=\frac{1}{4\pi }\oint\nolimits_{k}\oint\nolimits_{k^{\prime }}\frac{\mathbf{%
R}_{+}(k)-\mathbf{R}_{-}(k^{\prime })}{\left\vert \mathbf{R}_{+}(k)-\mathbf{R%
}_{-}(k^{\prime })\right\vert ^{3}}\cdot \lbrack \mathrm{d}\mathbf{R}%
_{+}(k)\times \mathrm{d}\mathbf{R}_{-}(k^{\prime })],  \label{L1}
\end{equation}%
which is a topological invariant \cite{Alexander}. Different links are
topologically inequivalent and cannot be continuously deformed into one
another without a topological phase transition.

In the Hermitian case ($\gamma =0$), the band energy is real and $F_{\pm
,z}\left( k\right) =0$; subsequently, $\theta _{\pm }=\pm \pi /2$ and $%
\left\vert r_{\pm }(k)\right\vert =r_{0}$. Both vector field curves are
located on surface of the torus. The orthogonality of eigenmodes in the
Hermitian system yields $r_{\pm }(k)=\pm r_{0}$. In the presence of
non-Hermiticity ($\gamma \neq 0$), the energy band is complex, and $\theta
_{\pm }\neq \pm \pi /2$. The minor radius for the vector field curve is $%
\left\vert r_{\pm }(k)\right\vert \neq r_{0}$, because of the
nonorthogonality of eigenmodes in the non-Hermitian system, and the two
vector field curves detach from the torus surface.

In Fig.~\ref{fig2}(a), the two vector field curves form a Solomon's knot
with the linking number $L=-2$. As non-Hermiticity increases, two
independent vector field curves meet at the exceptional points (EPs) \cite%
{Midya,Miri,Kawabata19L,Schomerus18,CTChan18X} and form a network. The EPs
are the nodes of the network with $\mathbf{F}_{+}\left( k\right) =\mathbf{F}%
_{-}\left( k\right) $ as a consequence of eigenmode coalescence. As
non-Hermiticity increases, the curves reform as two independent closed
loops; the vector field link changes through the reconnection associated
with the EPs, and the linking number of the vector field curves is reduced
by one, forming a Hopf link with linking number $L=-1$ [Fig.~\ref{fig2}(c)].
This exhibits the untying of the link topology through the anti-$\mathcal{PT}
$-symmetric coupling. After another EP occurs at even greater anti-$\mathcal{%
PT}$-symmetric coupling strength [Fig.~\ref{fig2}(d)], the two vector field
curves are unlinked [Fig.~\ref{fig2}(e)]. The linking number associated with
this phase is $L=0$ and the band is topologically trivial. Notably, the anti-%
$\mathcal{PT}$-symmetric coupling induces the topological phase transition
at the EP, which dramatically differs from the topological phase transitions
in other non-Hermitian systems with $\mathcal{PT}$-symmetric gain and loss 
\cite{JLPRA,Poli,Schomerus,Weimann,MPan,ZKL1905} as well as in non-Hermitian
systems with asymmetric coupling strengths \cite%
{ZGong,ZWang18,Kunst,LJinPRB,WHCPRB,YFChen20R}, where topological phase
transition is irrelevant to the EP. Counterintuitively, increasing the
dissipation can create nontrivial topology. More dissipation reduces the
strength of the anti-$\mathcal{PT}$-symmetric coupling, tying vector field
link or increasing the linking number instead of behaving destructively.

\begin{figure}[t]
\includegraphics[ bb=25 40 510 500, width=8.8 cm,clip]{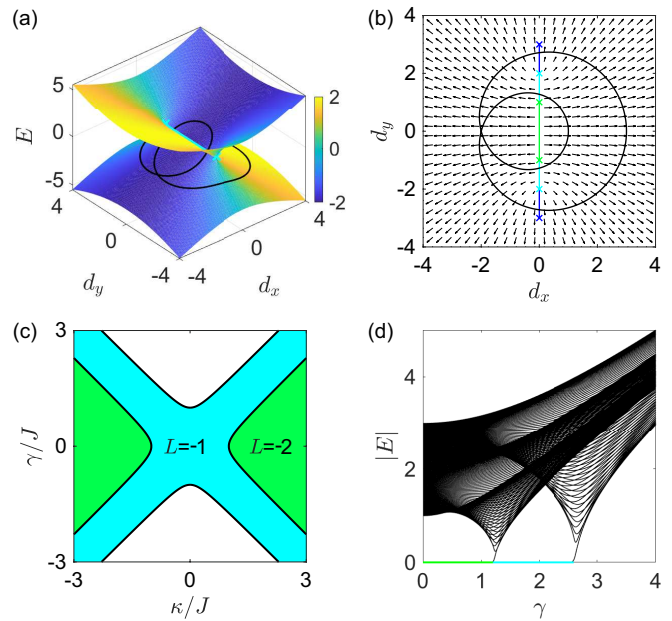}
\caption{(a) Complex spectrum of $h_{k}$ for $\vec{d}-i\vec{\Gamma}=(d_{x}-i%
\protect\gamma ,d_{y},0)$ at $\protect\gamma =2$. The color indicates the
imaginary part. The solid cyan line is the Fermi arc that connects two EPs
(crosses). The black curve is the effective magnetic field for the photonic
crystal lattice of Fig.~\protect\ref{fig1} at $J=1$, $\protect\kappa =2$;
its projection is shown at the top of (b) $(\mathbf{F}_{+,x},\mathbf{F}%
_{+,y})$. (c) Phase diagram. (d) Energy spectrum under open boundary.}
\label{fig3}
\end{figure}

The linking topology is closely related to the vortex associated with the
topological defects in the vector field. For the photonic crystal lattice
under consideration, the real part of the fictitious magnetic field
satisfies $d_{x}\left( k\right) =d_{x}\left( -k\right) $, $d_{y}\left(
k\right) =-d_{y}\left( -k\right) $, and $d_{z}\left( k\right) =0$. The types
of non-Hermiticity and their appearance manners are important for
non-Hermitian topological systems \cite{LJinPRB,ZKL1905}. If the
non-Hermiticity is $\vec{\Gamma}=\left( 0,0,\gamma \right) $, the
topological phase transition and the existence of the edge state are
unaltered because of the pseudo-anti-Hermiticity protection \cite%
{Esaki,ZKL1905}; the topological properties of the non-Hermitian system are
inherited by the EPs (exceptional rings or exceptional surfaces in 2D or 3D) 
\cite%
{ZhouH,Okugawa,Budich,Cerjan,LMDuan17L,QZhong19L,XFZhang19L,SHFan20B,Yamamoto19L,LLi}%
. If the non-Hermiticity is $\vec{\Gamma}=\left( 0,\gamma ,0\right) $, the
non-Hermitian skin effect occurs under open boundary condition \cite%
{TLee16L,ZGong,ZWang18,Kunst,ZWang19L,Longhi19R,SChen19B,HJZhang19B,CHLee19B,LJinPRB,WHCPRB,Longhi20L,YFChen20R,Sato20L,XZZhang20B,WXTeo,SPKou20B,SPKou20B2,SPKouIJPB20,ELee}%
, the non-Hermitian Aharonov-Bohm effect under periodical boundary condition
invalidates the conventional bulk-boundary correspondence \cite%
{LJinPRB,WHCPRB}, and the non-Bloch band theory is developed for topological
characterization \cite%
{Murakami19L,ZWang19L2,Borgnia,Kawabata20B,KawabataPRX,PXue20,Thomale20}.
Here, anti-$\mathcal{PT}$-symmetric coupling induces the imaginary part $%
\vec{\Gamma}=\left( \gamma ,0,0\right) $. $h_{k}$ lacks
pseudo-anti-Hermiticity protection. The inversion symmetry of $h_{k}$
ensures the absence of a non-Hermitian Aharonov-Bohm effect and the validity
of conventional bulk-boundary correspondence. The anti-$\mathcal{PT}$%
-symmetric coupling $\gamma $ in $h_{k}$ splits the diabolic point ($%
d_{x}=d_{y}=0$) in the Hermitian case ($\gamma =0$) into a pair of EPs [cyan
crosses in Fig.~\ref{fig3}(a)]. As $\gamma $ changes, the system switches
among topologically trivial phases and different topologically nontrivial
phases; therefore, the anti-$\mathcal{PT}$-symmetric coupling induces
topological phase transition and unties the vortex link.

Figure~\ref{fig3}(a) depicts the complex energy bands for $%
h_{k}=[d_{x}\left( k\right) -i\gamma ]\sigma _{x}+d_{y}\left( k\right)
\sigma _{y}$ as functions of $d_{x}$ and $d_{y}$ when $\gamma =2$ is fixed.
The Riemann surface spectrum has two EPs connected by the Fermi arc, where
the real part of the band gap closes. Two EPs located at $\left( 0,\pm
\gamma \right) $ in the $d_{x}$-$d_{y}$ plane are split from the diabolic
point [at the origin $\left( 0,0\right) $] for $\gamma =0$. The integer
topological charge associated with the diabolic point spits into two equal
half-integer topological charges in the presence of anti-$\mathcal{PT}$%
-symmetric coupling. The two-component planer vector field $(F_{+,x}\left(
k\right) ,F_{+,y}\left( k\right) )$ is depicted in Fig.~\ref{fig3}(b) \cite%
{VectorField}. The linking number of the vector field curves $\mathbf{F}%
_{\pm }\left( k\right) $ equals to the winding number of the planar vector
field, $w_{\pm }=\left( 2\pi \right) ^{-1}\int_{0}^{2\pi }\nabla _{k}\phi
_{\pm }\left( k\right) \mathrm{d}k=-L$, where $\phi _{\pm }(k)=\arctan
[F_{\pm ,y}\left( k\right) /F_{\pm ,x}\left( k\right) ]$. The half-integer
topological charge associated with EP is revealed from the argument of the
planar vector field $\mathbf{F}_{\pm }\left( k\right) $. The topological
charge associated with the diabolic point or EP is equal to the global Zak
phase \cite{Huang,SChen,Lieu,MPan}, which is the summation of the winding
numbers of the real fictitious magnetic field loop encircling each
individual EP. Notably, the global Zak phase can always be zero for an
inappropriate gauge of the eigenmodes \cite{Nesterov}, which is invalid for
topological characterization and not relevant to the winding number
associated with the gauge independent vector field for each band.

The topological phase transition occurs at the EPs, where the band gap
closes at exact zero energy. The topological phase transition points are 
\begin{equation}
\gamma ^{2}=\kappa ^{2}+J^{2}/2\pm \left( J/2\right) \sqrt{J^{2}+8\kappa ^{2}%
}.
\end{equation}%
The phase diagram in the $\kappa $-$\gamma $ plane is shown in Fig.~\ref%
{fig3}(c). The black curve in Fig.~\ref{fig3}(a) indicates one energy branch
of the photonic crystal lattice. The projection of the black curve in the $%
d_{x}$-$d_{y}$ plane is the real fictitious magnetic field loop $\vec{d}$
[Fig.~\ref{fig3}(b)], which exhibits time-reversal symmetry and mirror
symmetry with respect to the horizontal axis, $d_{x}\left( k\right)
=d_{x}\left( -k\right) $ and $d_{y}\left( k\right) =-d_{y}\left( -k\right) $%
. The real fictitious magnetic field loop $\vec{d}$ passes the EPs $\left(
0,\pm \gamma \right) $ at the gapless phase. In Fig.~\ref{fig3}(b), the
colored crosses indicate the EPs, and the colored lines indicate the Fermi
arcs at different $\gamma $. Each time the loop $\vec{d}$ comes across the
Fermi arc, the real part of the energy gap closes and two eigenmodes switch 
\cite{QZhong19NC,SHFan18B}. For weak non-Hermiticity, the real part of the
energy bands is gapped; the fictitious magnetic field loop passes {$%
d_{x}\left( k\right) =0$} four times as a result of the winding number $%
w_{\pm }=2$. Consequently, the magnetic field loop encircles the two EPs
(green crosses) twice, $F_{\pm }\left( k\right) $ form a Solomon's knot, and
the imaginary part of the energy band intersects four times [Fig.~\ref{fig2}%
(a)] at fixed momenta {$d_{x}\left( k\right) =0$} independent of
non-Hermiticity. As non-Hermiticity increases, the band gap diminishes; the
vector field loop passing {$\left( 0,\pm d_{y}\left( k_{0}\right) \right) $}
may cross the Fermi arc when {$\left\vert \gamma \right\vert >\left\vert
d_{y}\left( k_{0}\right) \right\vert $}. The crossing indicates that the
band gap of the real energy closes, and the real part instead of the
imaginary part of the energy band crosses at {$k=\pm k_{0}$}; this occurs
for large non-Hermiticity. We notice {two} and {four} crossings in the real
part of the energy bands in Figs.~\ref{fig2}(c) at $\gamma =2$ and~\ref{fig2}%
(e) at $\gamma =3$, respectively. In Fig.~\ref{fig2}(c), the magnetic field
loop encircles the two EPs (cyan crosses) once; in Fig.~\ref{fig2}(e), the
magnetic field loop does not encircle either of the two EPs (blue crosses).

\begin{figure}[tb]
\includegraphics[ bb=0 0 495 360, width=8.8 cm,clip]{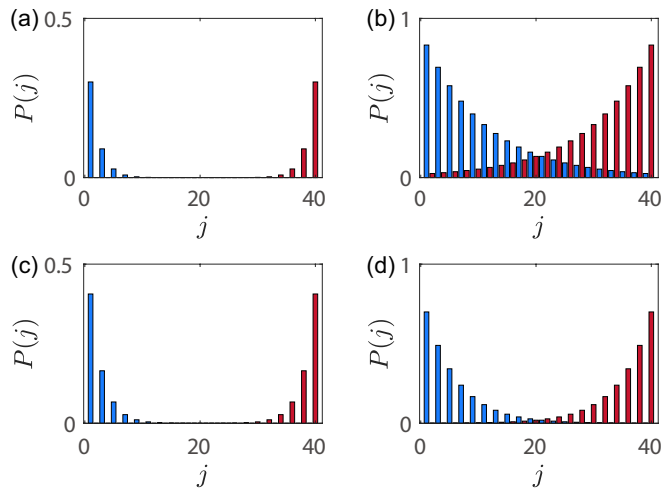}
\caption{Edge states in the topologically nontrivial phases. The parameters
are (a) $\protect\gamma =1$ with $\left\vert \protect\rho _{+}\right\vert <1$%
, (b) $\protect\gamma =1$ with $\left\vert \protect\rho _{-}\right\vert <1$,
(c) $\protect\gamma =1.274$, and (d) $\protect\gamma =2$. Other parameters
are $J=1$ and $\protect\kappa =2$. $P(j)=\left\langle \protect\psi %
_{j}\right\vert \left. \protect\psi _{j}\right\rangle $. The blue (red) bar
indicates the probability of the left (right) photonic edge modes on the
sublattice $A$ ($B$).}
\label{figES}
\end{figure}

\textit{Topological photonic zero modes.---}Topological phase transition is
inextricably related to the change of topological photonic edge modes.
Untying the link by $1$ reduces one pair of edge modes. Figure~\ref{fig1} is
a quasi-1D coupled Su-Schrieffer-Heeger chain; the termination under the
open boundary condition is indicated by the red lines, the open boundary
spectrum as a function of the anti-$\mathcal{PT}$-symmetric coupling
strength is depicted in Fig.~\ref{fig3}(d). The edge modes are zero modes,
which are analytically solvable at the limitation of infinite size, and
satisfy the steady-state equations of motion%
\begin{eqnarray}
0 &=&-i\gamma \psi _{A,j}+J\psi _{A,j+1}+\kappa \psi _{A,j+2}, \\
0 &=&-i\gamma \psi _{B,j}+J\psi _{B,j-1}+\kappa \psi _{B,j-2}.
\end{eqnarray}%
The edge modes decay to zero from one boundary to the other. The left
(right) edge modes localize in the sublattice $A$ ($B$) in blue (red). The
steady-state equations of motion are recurrence formulas. The localized edge
modes can be regarded as plane waves $e^{\pm ikj}$ with complex momenta $k$,
which are zeros of the complex function 
\begin{equation}
f\left( \rho \right) \equiv -i\gamma +J\rho +\kappa \rho ^{2}=0.
\end{equation}%
where $\rho =e^{\pm ik_{0}}=-(J\pm \sqrt{4i\kappa \gamma +J^{2}})/\left(
2\kappa \right) $ yields the localization length of the edge mode $\xi
=\left\vert \mathrm{Im}(k_{0})\right\vert ^{-1}$ \cite{Pollmann}. The number
of edge modes at either boundary is equal to the number of zeros $f\left(
\rho \right) =0$ within the region $\left\vert \rho \right\vert <1$, which
is identical to the linking number of the two vector field links. The
gapless phase is the boundary of two gapped phases, where topological phase
transition occurs. The number of topologically protected edge modes at the
gapless phase is equal to that at the gapped phase with a small winding
number; two (zero) topological edge modes present in the gapless phase
between topological phases $L=-2$ ($L=0$) and $L=-1$. The left edge mode
amplitude is $\psi _{A,j}=\rho ^{j}$ and the right edge mode amplitude is $%
\psi _{B,N+1-j}=\rho ^{j}$. The two pairs of edge modes at $\gamma =1$ are
depicted in Figs.~\ref{figES} (a) and~\ref{figES}(b), respectively. The pair
of edge modes at $\gamma =1.274$ and $\gamma =2$ are depicted in Figs.~\ref%
{figES}(c) and~\ref{figES}(d).

\textit{Conclusion.}---We investigate a dissipative quasi-1D topological
photonic crystal lattice of coupled resonators. Counterintuitively, more
dissipation constructively generate nontrivial topology from trivial
topology. The coupled resonator array exhibits sublattice symmetry with
alternatively presented anti-$\mathcal{PT}$-symmetric coupling. The
nontrivial topology of the two-band lattice is established from the linking
geometry of the vector fields, which are the average values of the Pauli
matrices under the eigenmodes of the Bloch Hamiltonian. The anti-$\mathcal{PT%
}$-symmetric coupling alters the band topology and induces topological phase
transition at the EPs, associated with the untying of vector field links.
The linking number determines the number of topologically protected photonic
zero modes. Our findings provide novel insight for the influence of
non-Hermitian anti-$\mathcal{PT}$-symmetric coupling in the topological
photonics and deepen the understanding of 1D topology in non-Hermitian
systems.

\acknowledgments This work was supported by National Natural Science
Foundation of China (Grants No.~11975128 and No.~11874225).

\end{document}